# AN EXAUSTIVE SURVEY OF TRUST MODELS IN P2P NETWORK

S. Udhaya Shree[1] and Dr. M. S. Saleem Basha[2]

[1]Department of Computer Applications, Rajiv Gandhi College of Engineering and Technology, Puducherry, India
[2]Department of Computer Science, Mazoon University College, Oman

## ABSTRACT

*Most of the peers accessing the services are under the assumption that the service accessed in a P2P network is utmost secured. By means of prevailing hard security mechanisms, security goals like authentication, authorization, privacy, non repudiation of services and other hard security issues are resolved. But these mechanisms fail to provide soft security. An exhaustive survey of existing trust and reputation models in P2P network regarding service provisioning is presented and challenges are listed. Trust issues like trust bootstrapping, trust evidence procurement, trust assessment, trust interaction outcome evaluation and other trust based classification of peer's behavior into trusted,, inconsistent, un trusted, malicious, betraying, redemptive are discussed,*

## 1. INTRODUCTION

In PEER-TO-PEER (P2P) systems peers collaborate knowingly or unknowingly with other peers in the network for the sake of accomplishing the tasks. Hence there exist a large scale security threat for P2P systems. Security is a term that is being renowned by the research community and to narrow down the focus on security the difference between "soft security" and "hard security" was first coined by Rasmusson and Jansson[21] who used the term hard security for traditional mechanisms like authentication and access control, and soft security for social control mechanisms. Soft security tries to control the social security threats and avoids social conflicts. By the creation of long-term trust relationships among peers, the network can provide a more secured environment, there by reducing risk and uncertainty in future interactions among peers. However, the establishment of trust relation between the peers involved in the interactions is difficult in such a malicious open system. Trust is a social concept and hard to measure with numerical values. In the literature trust and reputation are interchangeably used. "But trust is a complex concept and in many cases the definition for trust is measured in terms of confidence that an entity of a system places on another entity of the same system for performing a given task with main focus on two features namely uncertainty and subjectivity whereas reputation is a more objective concept, and is based on information about or observations of the past behaviour of an entity. Both concepts are very related, and in fact, reputation can be used as a means to determine whether an entity can trust another entity" [22]

*Benefits of application of trust and soft security include the following :

- More Trusted customer service ,
- Build trust relationship between trading partners,
- Effective use of technologies,
- Providing soft security and
- Increase companies' reputation.

This survey provides an exhaustive study of existing trust models by means of literature survey in section 2   security risks in section 3   analysis of trust models in section 4   and conclusion and future work  in section 5 and references in section 6.

## 2. RELATED WORK

There is a continuous effort in the research community to explore the challenges of trust and reputation models. This can be seen from the existing trust models. Still there are several issues or challenges yet to be tackled which can be seen from the following survey on trust and reputations models as presented in the literature. The table 2.1 depicts the different trust evaluation models in the P2P or multi agent environment. This table provides   trust evaluation models in the following order:

*1-Cuboid trust, 2- Eigen trust, 3- BNBTM,   4 – GroupRep,     5- AntRep, 6 - Semantic Web, 7- Global Trust, 8- Peer Trust,     9- PATROL – F, 10 – Trust evolution, 11- TDTM, 12- TACS,    13- SORT.*

In this article, we have studied several trust and reputation models and issues such as   trust bootstrapping, trust evidence, trust assessment, second order issues, interaction outcome evaluation, punishment, reputation propagation, redemption, context awareness, rewarding, dynamic nature and  trust type value  are being analyzed. Trust bootstrapping deals with the initial trust value assignment which is the value a truster assigns to trustee. 'First impression is the best impression' and a  wrong judgment results in bad transaction result. Trust evidence can involve direct or indirect interaction between truster and trustee. Second order  issues are security threats that are prevailing in the P2P as well as other network environments, namely individual malicious person attack or group of persons with bad intention,  collusion attack, sybil attack or impersonation, camafluage attack or on/off attack, trusted peer changing nature etc.There have been several solutions for each of these attacks.  In some of the existing trust models wide coverage of all attacks not being carried out. Trust interaction   evaluation is done by watch dog, centralized or de centralized node, Public key infrastructure (PKI), monitoring node, etc.  The trust evaluation may be    performed locally or globally. In certain models, good service or transaction is rewarded by providing weightage to the satisfaction factor or if trust level crosses threshold value. Diminishing effect deals with trust decay over a period. In trust models it has become necessary  to include context awareness. Trust evolves over a period of time, hence the trust model should be  a dynamic one. The trust value can be discrete or continuous, but continuous trust value is preferable over the discrete.  Survey papers taken into account are described below   as in the same order given in Table 2.1.

## 1. CuboidTrust

CuboidTrust[1] is a global trust model for p2p networks. It denotes the reputation which represents peer's trustworthiness by four relations. A cuboid is created by using coordinates (x,y,z) where z – quality of resource/file, y – peer that requested the resource and x – the peer who has given the feedback about the resource and denoted by *Px,y,z*. Binary rating is used global trust for each peer is calculated using power iteration of all the values stored by the peers [1].

## 2. EigenTrust

EigenTrust is a global trust model in a P2P, dealing with file sharing. Local trust is computed by the satisfactory rate of file downloading is defined as $Sij = sat(i,j) - unsat(i,j)$, where *sat(i,j) denotes the file* downloads by *i* from *j* and *unsat(i,j)* is the unsatisfactory downloads. The Global trust is can be obtained from the Power iteration formula [2].

Table 2.1 Trust Evaluation models on multiple issues

| Issues | Paper ID / Measures | 1 | 2 | 3 | 4 | 5 | 6 | 7 | 8 | 9 | 10 | 11 | 12 | 13 |
|---|---|---|---|---|---|---|---|---|---|---|---|---|---|---|
| Trust Bootstrapping (5) | Low value | | | | | | | | | | | | | x |
| | Neutral value | | | | | | | | | | | | | |
| | Pre-trusted | | x | | | | | | | | | | | x |
| | No. of interactions | | | | x | | | | | | | | | x |
| | Recent interactions | | | | | | | | | | | | | x |
| | Subtotal | | 1 | | 1 | | | | | | | | | 4 |
| Trust Evidence (5) | Direct | x | x | x | | x | | x | x | x | x | x | x | x |
| | Indirect (Reputation) | x | | x | x | | x | x | x | x | | x | x | x |
| | Indirect (Recommendations) | | x | | | | | x | x | | x | x | x | x |
| | Indirect (Referrals) | x | | | | | | | | | | | | |
| | Role | | | | | | | | | | | | | |
| | Subtotal | 3 | 2 | 2 | 1 | 1 | 1 | 3 | 3 | 2 | 2 | 3 | 3 | 3 |
| Trust Assessment (18) | Recent | | | | | | | | x | | x | x | x | x |
| | Size | | | | | | | | x | | x | x | x | x |
| | Category | | | | | | | | x | | | | | |

International Journal on Web Service Computing (IJWSC), Vol.5, No.3, September 2014| | | | | | | | | | | | | | | |
|---|---|---|---|---|---|---|---|---|---|---|---|---|---|---|
| | History | x | x | x | | x | | | x | | | | x | x |
| | Satisfaction | x | x | x | | x | | | x | | x | | x | x |
| | Adaptive | | | | | | | | | | | | x | |
| | Credibility | x | | | | | | | x | x | x | | | x |
| | Risk tolerance | | | x | | | | | | | | | | |
| | Similarity | | | | | | | | | | | | | |
| | Role based | | | | | | | | | x | | | | |
| | Sudden deviation | | | | | | | | | | | | | x |
| | Trust decay | | | | | x | | | | x | | | x | x |
| | Community trust | | x | | | | | | | | | | | |
| | Local | x | x | x | x | x | x | x | x | x | | | x | |
| | Global | x | x | x | | x | x | x | x | x | | | x | |
| | Confidence as ser. Provider | | | | | | | | | | | | | x |
| | Confidence as recommend-er | | | | | x | | | | | | | | |
| | Different weights for recent | | | | | | | | x | | | | | |
| | Subtotal | 5 | 5 | 5 | 1 | 5 | 3 | 2 | 9 | 5 | 4 | 2 | 8 | 8 |
| Second order issues(9) | Individual malicious | | | | | | | | | | | | | x |
| | Collusion | x | x | | | | | | x | | | | | |
| | Sybil attack | | x | | | | | | | | | | | x |
| | On-off attack | | | | | | | | | | | | | x |
| | Ballot stuffing | | | | | | | | | | | | | x |
| | Man in the middle | | | | | | | | | | | | | |
| | Bad mouthing | | | | | | | | | | | | | x |
| | Partially malicious | | | | | | | | | | | | | x |
| | Malicious pre-trusted peers | | | | | | | | | | | | | |
| | Subtotal | 1 | 2 | | | | | | 1 | | | | | 6 |
| Interaction outcome evaluation (5) | Watch dog | | | | | | | | | | | | | |
| | PKI | | | | | | | | | | x | | | |
| | Beacon node | | | | | | | | | | | | | |
| | Mobile agents | | | | | | | | | | | | | |
| | Node itself | x | x | x | | | | | x | | | | | x |
| | Subtotal | 1 | 1 | 1 | | | | | 1 | | 1 | | | 1 |
| Reputation propagation(4) | Alarm message | | | | | | | | | | | | | |
| | Positive Reputation value | | | | | | | | | | | | x | |
| | One hop neighbours | | | | | | | | | | | | | x |
| | Agents | | | | | | | | | | | | | |
| | Subtotal | | | | | | | | | | | | 1 | 1 |
| Punishment(4) | Low reputation value | | | | | | | | | | | | | |

| | | | | | | | | | | | | | |
|---|---|---|---|---|---|---|---|---|---|---|---|---|---|
| | Threshold value | | | | x | | | | | | | x | |
| | Time elapse | | | | | | | | | | | | |
| | Forwarding Service refusal | | | | | | | | | | | | |
| | Subtotal | | | | 1 | | | | | | | 1 | |
| Redemption(2) | Opportunity given | | | | | | | | | | | | |
| | Consistency of co-operation | | | | | | | | | | | | |
| | Subtotal | | | | | | | | | | | | |
| Context aware(1) | | | | | | | | | | | | | |
| Rewarding(2) | Satisfaction | x | | | | | x | | | | x | | |
| | Threshold | | x | | x | | | | | | | | |
| | Subtotal | 1 | | 1 | 1 | | 1 | | | | 1 | | |
| Dynamic(1) | | | | | x | | x | | | | | | |
| | Subtotal | | | | 1 | | 1 | | | | | | |
| Trust value type(2) | Discrete | x | x | x | | | | | | | | | |
| | Continuous | | | | x | | | | | | | | x |
| | | 1 | 1 | 1 | 1 | | 1 | | | | | | 1 |
| | Total trust elements | 12 | 12 | 10 | 3 | 10 | 4 | 5 | 16 | 8 | 6 | 6 | 14 | 24 |

## 3. BAYESIAN NETWORK BASED TRUST MANAGEMENT (BNBTM)

BNBTM considers multiple features of an applications to denote the trust in various factors and evaluates by a single Bayesian network. Beta probability distribution functions uses past experiences to evaluate the trust [3].

## 4. GROUPREP

GroupRep is representing the trust among group members. This includes three levels of trust namely, trust between groups, trust developed between groups and peer trust for another peer [4].

## 5. ANTREP

AntRep algorithm is based on bio inspired swarm intelligence algorithm. Every peer maintains a reputation table giving reputation of 'n' number of peers in the network. The reputation table slightly differs from the routing table in the sense that (i) instead of distance between peers reputation of the peer is stored; (ii) The reputation value is used as the metric for the selection of peer. Two types of ants used namely forward ants and backward ants are used for finding reputation values of peers and to propagate these reputation value over the network. Initially from the reputation table a peer with the highest reputation value is selected a unicast ant is sent to that peer for transaction. If no such highest value exist in the table then broadcast ants are sent along all the paths [5]. After the transaction is over, a backward ant is used to update all the reputation values of all the nodes/peer on its way.

## 6. SEMANTIC WEB

Zhang et al.[6], have presented a trust model for multi agent system. The final trust value on the path connecting two agents is assigned by adding the trust of individual edges multiplied by corresponding weights associated with each edge.

## 7. GLOBAL TRUST

Instead of concentrating local trust value of a node, by accumulating the local trust values, the global trust value of a node is evaluated as given in [5,6,7]

## 8. PEER TRUST

This work is a reputation-based trust model. Based on three factors namely number of transactions, credibility of the peer and the feedback a peer receives from other peers to calculate the adaptive trust.[8].

## 9. PATROL-F (comPrehensive reputAtion-based TRust mOdeL- Fuzzy)

PATROL-F includes many important concepts direct experiences, reputation values, credibility, time based decay of information, first impressions and a node system hierarchy for the purpose of computing peer reputation. This model uses fuzzy logic for the categorizing the peer based on trust level into *"good"* or *"better"* and *"bad"* or *"worse"* [9].

## 10. TRUST EVOLUTION

Wang et al., have developed a trust model for P2P networks. It uses direct trust and recommendation from other peers and also considers context and trust lies within the interval [0,1] [10].

## 11. TIME-BASED DYNAMIC TRUST MODEL (TDTM)

TDTM is a bio inspired technique using ant colony algorithm that represents trust between the nodes as the pheromone value on the edge connecting the two nodes in the network. [1].

## 12. TRUST ANT COLONY SYSTEM (TACS)

TACS is based on the ant colony system.. In this model the most trustworthy node is selected for service request based on the pheromone traces on the path. Every link is associated with pheromone value representing the trust one peer has over the other. Ants travel along every path depositing pheromone and finds the most trustworthy path leading to the most reputable server [12].

## 13. SORT

In SORT[13], to evaluate interactions and recommendations in a better way, importance, recentness, peer satisfaction, recommender's trustworthiness and confidence about recommendation are considered. Additionally, service and recommendation contexts are separated. Taking into account real-life factors, simulations are carried out more realistically.

## 14. PATROL (comPrehensive reputAtion-based TRust mOdeL)

PATROL is a reputation based trust model for distributed computing, considering multiple factors such as reputation values, direct experiences, credibility, time based trust, first impressions, similarity, popularity, activity, cooperation between hosts, role based trust consistency and confidence. PATROL takes into account different weightage for different factors. [14].

## 15. META-TACS

META-TACS is an extension of the TACS algorithm developed by the Felix et.al. [12]. They have extended the TACS model by optimizing the working parameters of the algorithm using genetic algorithms [15].

## 3. SECURITY RISKS

In an open network, individuals or peers are to be identified as benevolent peers or malicious one based on the trust value. There is a possibility of change in behaviors of an individual. These behavioral changes are subjective in nature.

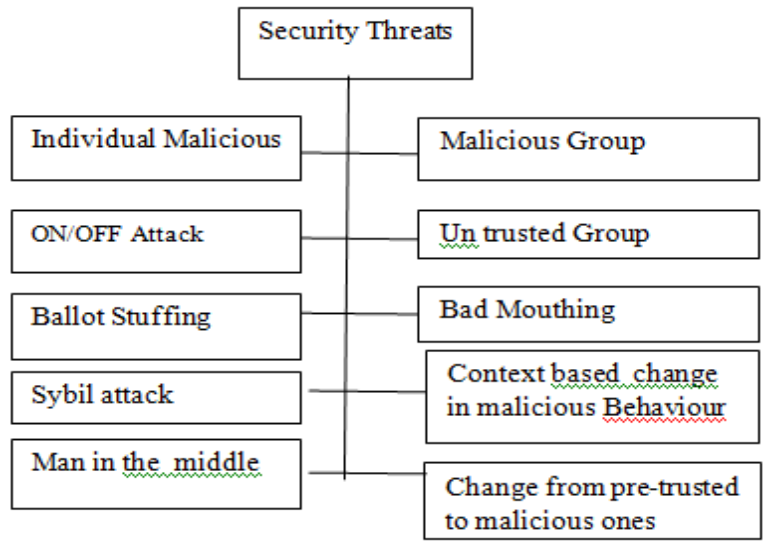

Figure 1. Security Threats

The above mentioned risks are to be taken care by the trust and reputation models and there should be comprehensive model to identify, mitigate, provide and recover from all types of attacks .In the existing works only some of the issues are provided with a solution and demands some additional effort to accomplish a more secured environment

## 4. ANALYSIS OF TRUST MODELS

From the survey, it is known that only 25 % of trust based issues or challenges have been covered in current P2P network.

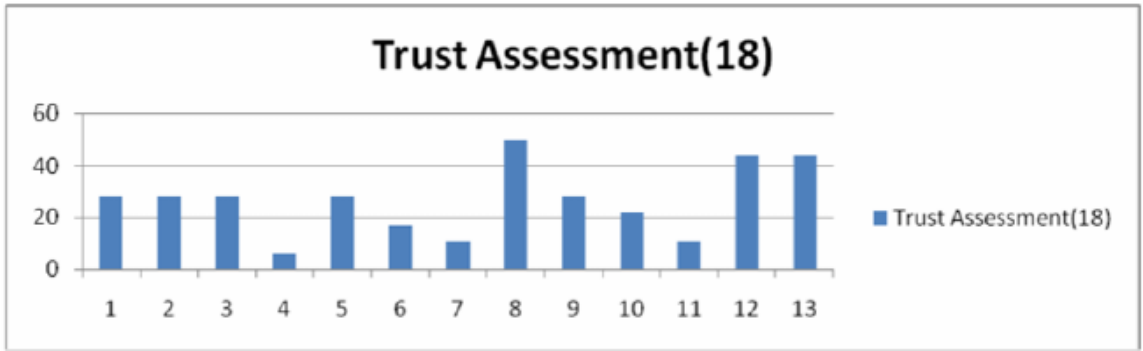

Figure. 2. Trust influencing factors

In the above graph paper-id is given on the X axis and the coverage trust assessment factors in each of the papers is given along the Y axis. From the survey it can be seen that trust can be assessed by 18 different attributes of an entity or environment. Only 50 % of the trust sources are utilised for the assessment while there still 50% sources that are yet to be tackled. Risk tolerance, Similarity, role based trust, sudden behaviour change, trust decay communityRisk tolerance nature , Similarity among the peers, Role played by the peer, Sudden deviation in bevaiour, Trust decay, community based trust, confidence as provider and requetor etc. are the other elements which should be given due weightage while computing trust. In order to arrive at a more comprehensive trust and reputation model, some more attempts have to be taken for establishing an effective trustworthynvironment in P2P network.

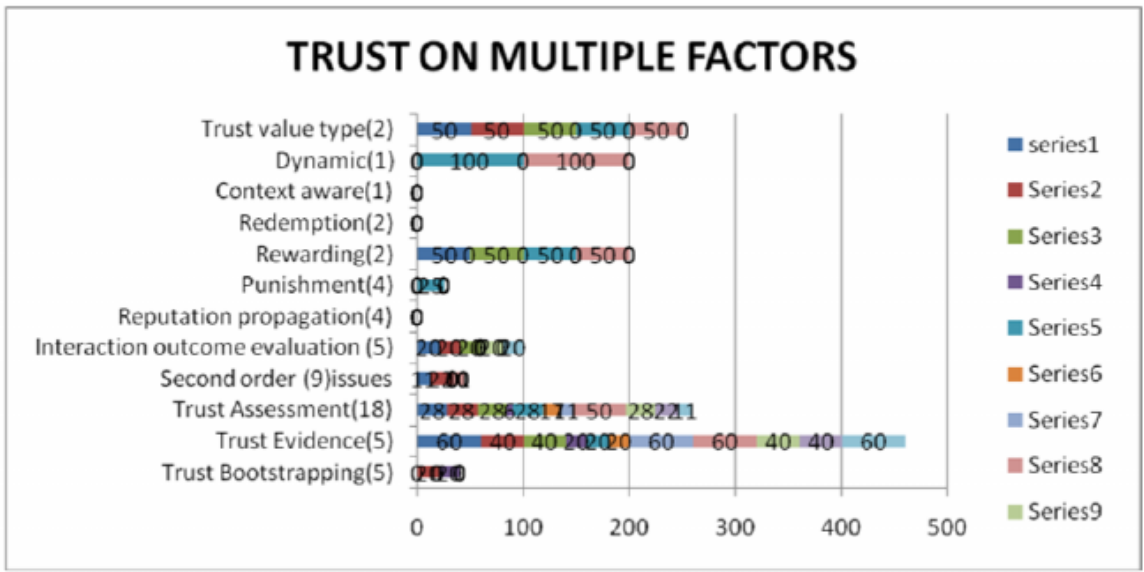

Figure 3. Trust on multiple factors

In the above figure 3, the main issues in the trust and reputation model explored in various articles has been given along the X axis and the issues along the Y axis. It can be seen from the above graph that trust factors like context awareness, redemption, reputation propagation have not been tackled in many of the trust and reputation models. Context depicts the environment. Different situations results in different behaviours of the peers. Consistent good behviour should be given more weightage. A chance should also be given for a malicious person to become a benovalent one. Some means of reputation propagation should be encouraged to identify the trusted group. But at the same time measures should be taken to curb the badmouthing peers. So understanding soft trust based attacks helps a peer to be more vigilant and continue to leverage the available services in the network. Hence the second order isssues have been analysed in the above models and figure 4. Presents the issues by means of pie chart. The graph given in figure 4. shows the coverage of various security attacks explored in different articles as depicted in figure 1. and it is known that issues like are not being tackled in all the trust models taken in literature survey. Hence, these issues shoule be effectively handled in the forthcoming trust models to provide a smooth trust worthy transactions or interactions in P2P networking environment. It can be seen that SORT[13], has covered 78 % of trust issues paving way to explore further in this direction.

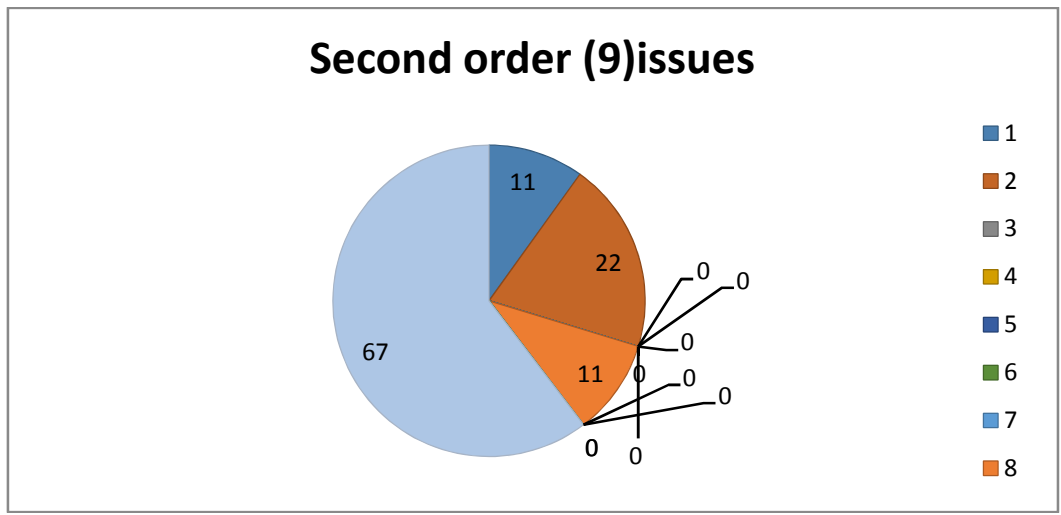

Figure 4. Second order issues

## 5. CONCLUSION AND FUTURE WORK

P2P network provides an efficient means of data communication. In this survey, the trust issues explored in the existing trust and reputation models have been analyzed. Our intention is to provide and execute a trustworthy P2P model. We emphasis that solution to multiple soft security based threats should be given more effectively taking into account multifaceted approach and trust mechanism entrust a healthy and smooth data transfer and services between peers. It is known that from figure 3. Out of 18 trust issues, context awareness, redemption, reputation propagation, second order issues and trust bootstrapping are the areas one has to perform intensive exploration considering trust as a subjective trust and must resolve with dynamic and innovative solutions. The survey paper SORT[13] covers 67% of trust issues but it has started the trust bootstrapping process with low value and pre trusted peer's value But in the case of no pre trusted peers available, this model cannot solve the bootstrapping issue. In case of trust assessment it has not categorized the peers into similar peers or role based peers, local, global, community trusted peers and adaptively is missing. When considering the indirect trust experience, referrals are not taken into account and weightage for confidence as recommender is not used. While considering second order issues collusion attack, man in the middle attack, pre trusted peers changing into malicious category are not being explored. When interaction outcome evaluation is done by the node itself, there is possibility of misjudgment. It also does not cover the trust decay and punishment activity. Changing into benevolent one is not being rewarded and hence dynamic and context awareness factors are missing. Hence in our future work we like to provide an efficient dynamic trust worthy framework for service provisioning and leveraging taking into account the subjective nature of trust and giving much importance for the issues like bootstrapping, redemption, context awareness and reinforcement.

## Authors

S. Udhaya Shree is working as Assistant Professor in the Department of Computer Applications, Rajiv Gandhi College of Engineering, Pondicherry. Currently, she is pursuing Ph.D. in Computer Science and Engineering at Pondicherry University, Pondicherry. She has obtained M.Sc. (Maths) and M.C.A. degree from Madras University, Chennai, India. She has done M.Tech. in Computer Science and Engineering from Manomanium Sundaranar University, Tirunelveli, India. Currently she is pursuing Ph.D. in Computer Science and Engineering at Pondicherry University, Puducherry, under the guidance of Dr. Saleem Basha M.S. 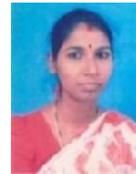 Her areas of interest are formal verification of QoS of Web Services using Timed Automata, Web Service Composition using Bio inspired Optimization Techniques. She has published more than 12 papers in National and International Conferences.

Dr. Saleem Basha.M.S is working as Assistant Professor & Research Director in the Department of Computer Science, Mazoon University College, Muscat, Sultanate of Oman. He has obtained B.E in the field of Electrical and Electronics Engineering, Bangalore University, Bangalore, India and M.E in the field of Computer Science and Engineering, Anna University, Chennai, India and Ph.D. in the field of Computer Science and Engineering in Pondicherry University, India. He is currently working in the area of Hackers psychology, SDLC specific effort estimation models and web service modeling 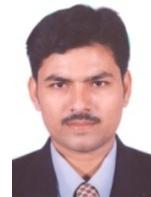 systems. He has published more than 70 research papers in National and International journals and conferences.